\documentclass[%
 aip,
 amsmath,amssymb,
 reprint,%
]{revtex4-1}

\usepackage{hyperref}
\usepackage{graphicx}
\usepackage{dcolumn}
\usepackage{bm}

\usepackage[utf8]{inputenc}
\usepackage[T1]{fontenc}
\usepackage{mathptmx}
\usepackage{etoolbox}
\usepackage{siunitx}

\usepackage{xcolor}
\newcommand{\HT}[1]{\textcolor{black}{#1}} 
\newcommand{\AI}[1]{\textcolor{black}{#1}} 

\makeatletter
\def\@email#1#2{%
 \endgroup
 \patchcmd{\titleblock@produce}
  {\frontmatter@RRAPformat}
  {\frontmatter@RRAPformat{\produce@RRAP{*Authors to whom the correspondence should be addressed: {#2}}}\frontmatter@RRAPformat}
  {}{}
}%
\makeatother
\begin{document}
\preprint{AIP/123-QED}

\title{Coupled vertical double quantum dots at single-hole occupancy}

\author{Alexander S. Ivlev}
\altaffiliation{These authors contributed equally}
\affiliation{QuTech and Kavli Institute of Nanoscience, Delft University of Technology, PO Box 5046, 2600 GA Delft, The Netherlands}
\email{A.S.Ivlev, a.s.ivlev@tudelft.nl; M.Veldhorst, m.veldhorst@tudelft.nl}
\author{Hanifa Tidjani}
\altaffiliation{These authors contributed equally}
\affiliation{QuTech and Kavli Institute of Nanoscience, Delft University of Technology, PO Box 5046, 2600 GA Delft, The Netherlands}
\author{Stefan D. Oosterhout}
\affiliation{QuTech and Netherlands Organisation for Applied Scientific Research (TNO), Delft, The Netherlands}
\author{Amir Sammak}
\affiliation{QuTech and Netherlands Organisation for Applied Scientific Research (TNO), Delft, The Netherlands}
\author{Giordano Scappucci}
\affiliation{QuTech and Kavli Institute of Nanoscience, Delft University of Technology, PO Box 5046, 2600 GA Delft, The Netherlands}
\author{Menno Veldhorst}
\affiliation{QuTech and Kavli Institute of Nanoscience, Delft University of Technology, PO Box 5046, 2600 GA Delft, The Netherlands}

\date{15 January 2023}

\begin{abstract}
 
    Gate-defined quantum dots define an attractive platform for quantum computation and have been used to confine individual charges in a planar array. Here, we demonstrate control over vertical double quantum dots confined in a \AI{strained germanium double quantum well}. We sense individual charge transitions with a single-hole transistor. The vertical separation between the quantum wells provides a sufficient difference in capacitive coupling to distinguish quantum dots located in the top and bottom quantum well. Tuning the vertical double quantum dot to the (1,1) charge state confines a single hole in each quantum well beneath a single plunger gate. By simultaneously accumulating holes under two neighbouring plunger gates, we are able to tune to the (1,1,1,1) charge state. These results motivate quantum dot systems that exploit the third dimension, \AI{creating} opportunities for quantum simulation and quantum computing.

\end{abstract}

\maketitle

\begin{figure*}
    \centering
    \includegraphics[width=\textwidth]{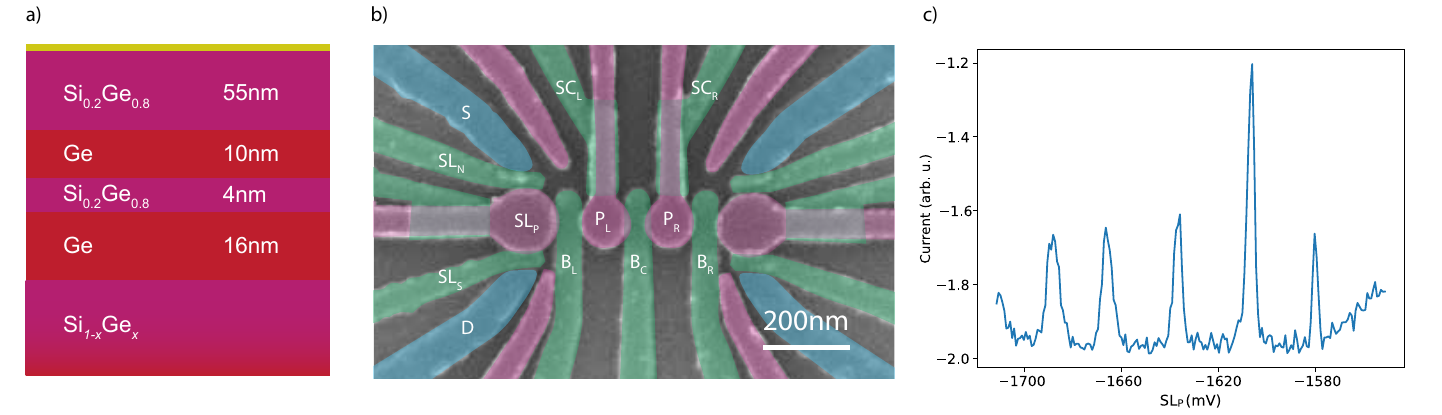}
    \caption{\textbf{Double quantum well heterostructure and top gate layout.} \textbf{a} Schematic of the double quantum well heterostructure \AI{with the numbers indicating the targeted layer thickness}. The yellow layer denotes the \AI{native} SiOx. \textbf{b} False coloured SEM of a device nominally identical to the one used in this experiment. The \AI{left-most plunger gate acts as a charge sensor and the two central plunger gates and surrounding barrier gates confine individual holes under P\textsubscript{L,R}. The remaining right side of the device forms a hole-reservoir.} \textbf{c} Typical Coulomb oscillations of the single-hole transistor formed underneath the plunger gate SL\textsubscript{P}, at a typical source-drain bias of \SI{100}{\micro V}.}
    \label{fig:fig1_Device}
\end{figure*}
Attaining control over individual charges in silicon\cite{Angus_NanoLetters_2007,Simmons_APL_2007_SiliconQD,Zwanenburg_2013_Silicon_RevModPhys} and germanium \cite{Hendrickx_Nature_2018_GermaniumQD,Lawrie_2020_APL_GeQD,Scappucci_2021_NatureRevMat_GermaniumRoute} constituted a necessary prerequisite to enable quantum computation with gate-defined quantum dots\cite{Loss_PRA_1998_SpinQubit,Burkard_2023_RevModPhys_SemiconQubits}. Planar quantum dot systems have progressed significantly, supporting high-fidelity single and two-qubit logic, multi-qubit logic, rudimentary error correction, and control over a 16 quantum dot array\cite{Yoneda_2018_Nature_HighFidelity, Lawrie_NatureComm_2023_SimultaneousControl, Hendrickx_2021_Nature_4qubit, Philips_2022_Nature_SixQubitProcessor,Yoneda_2018_Nature_HighFidelity,Xue_2022_Nature_FidelityTheshold,Noiri_2022_Nature_FidelityThreshold,Takeda_2022_Nature_QEC,VanRiggelen_2022_NatureQuantum_QEC, Borsoi_2023_Nature_16DotArray}. The development of a double germanium quantum well heterostructure \cite{Tosato_2022_AdvQuantTech_DoubleQuantumWell} has enabled the realisation of a vertically coupled double quantum dot \cite{Tidjani_2023_PRApplied_VerticalDQD}, by taking advantage of the third dimension. Gaining control over single charges confined in quantum dots in multilayer systems may become a key asset in obtaining high connectivity in large quantum dot arrays \cite{Tidjani_2023_PRApplied_VerticalDQD}. In the near term, single-charge control in bilayer quantum dot systems may enable the realization of small-scale quantum simulators of magnetic phases in correlated spin systems~\cite{Buterakos_2023_PRB_BilayerQSimulator}.\\

Here, we demonstrate a vertical double quantum dot formed under a single plunger gate and tuned to single-hole occupancy. The occupancy is detected by charge sensing with a single-hole transistor. Using a second plunger gate, the system is extended to a \AI{vertical} 2x2 quantum dot array in the x-z plane parallel to the (100) heterostructure growth direction, filled down to the (1, 1, 1, 1) hole occupation. In comparison, achieving such a charge configuration in planar systems is non-trivial and have been demonstrated only recently in planar germanium \cite{VanRiggelen_2021_APL_SingleHole2D} and silicon \cite{unseld_2023_APL_2x2Silicon}. \\


Fig.~\ref{fig:fig1_Device}a depicts a schematic of the Ge/SiGe heterostructure, grown by reduced pressure chemical vapor deposition as detailed in Tosato et al.\cite{Tosato_2022_AdvQuantTech_DoubleQuantumWell}. The heterostructure features two strained Ge quantum wells \AI{with thicknesses of 16 nm and 10 nm embedded in strain-relaxed Si\textsubscript{0.2}Ge\textsubscript{0.8}}. The separation between the quantum wells is 4 nm and the separation of the top quantum well from the semiconductor-dielectric interface is 55 nm, \AI{in line with} current heterostructures\cite{Lodari_2021_MQT_Heterostructure} hosting spin qubit devices. Ti/Pd metallic gates (Fig.~\ref{fig:fig1_Device}b) are fabricated in two layers and separated by Al\textsubscript{2}O\textsubscript{3}, to electrostatically confine holes in the quantum wells (for further details on fabrication see~\cite{Tidjani_2023_PRApplied_VerticalDQD}). Four plunger gates are patterned\AI{, with the left-most plunger gate SL\textsubscript{P} forming a charge sensor and the right-most acting only as a reservoir in this experiment}. The barrier gates SL\textsubscript{N (S)} control the tunnelling between the charge sensor and the ohmic contacts. We define quantum dots localised in the two quantum wells using plunger gates P\textsubscript{L} and P\textsubscript{R}, and barrier gates B\textsubscript{L}, B\textsubscript{C} and B\textsubscript{R}. Additionally, screening gates SC\textsubscript{L} and SC\textsubscript{R} provide further fine-tuning and prevent the formation of unwanted quantum dots. Barrier gates B\textsubscript{L} and B\textsubscript{R} also control the loading of charge carriers from the reservoirs to the quantum dots. 

To facilitate charge sensing, a \SI{100}{\micro V} bias is applied across the ohmic contacts S and D. The current signal through the sensor is determined by two-terminal DC measurements using low impedance lines and resulting in an integration time in the order of \SI{100}{\micro s}. We calibrate the gate voltages to observe well-defined Coulomb peaks corresponding to the transport of holes through the single hole transistor (SL\textsubscript{P}), as seen in Fig. \ref{fig:fig1_Device}c. At the edge of a Coulomb peak, the source-drain current is highly sensitive to the electrostatic environment and in particular to the charge occupation of any quantum dots under plunger gates P\textsubscript{L} and P\textsubscript{R}, similar to charge sensors in single quantum well systems. During all following measurements the voltage on SL\textsubscript{P} is tuned such that it maintains a high sensitivity to the studied charge states. Previous works have observed that the transport signal through a single-hole transistor may be diminished in a double quantum dot regime~\cite{Tidjani_2023_PRApplied_VerticalDQD}, therefore we carefully tune the sensor to obtain regular and well-defined Coulomb peaks. We speculate that in this regime only one quantum well is contributing to transport through the charge sensor \AI{(see Supplementary II)}. 

\begin{figure*}
    \centering
    \includegraphics[width=\textwidth]{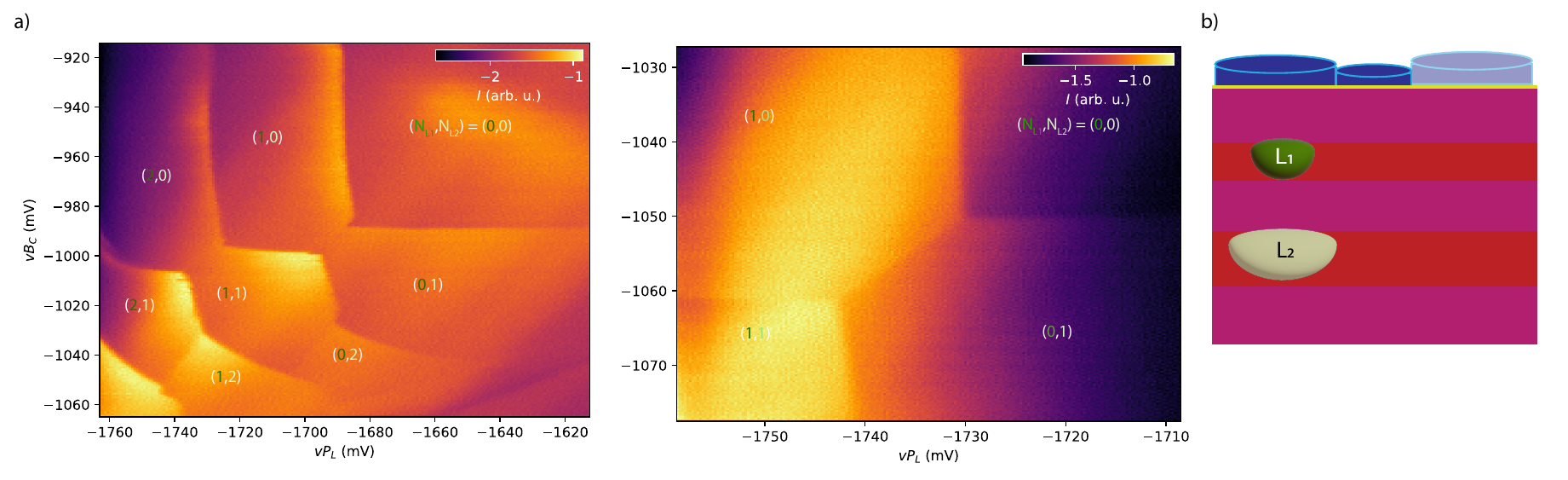}
    \caption{\textbf{Single-hole occupancy in a vertical double quantum dot.} \textbf{a.} The left panel shows the charge-stability diagram of a double quantum dot formed underneath plunger gate P\textsubscript{L} measured by charge sensing. The occupation (N\textsubscript{L1},N\textsubscript{L2}) for quantum dots QD\textsubscript{L1} and QD\textsubscript{L2} is noted in each region and is controlled by the gate voltages on P\textsubscript{L} and B\textsubscript{C}, which are applied as virtual gates vP\textsubscript{L}=P\textsubscript{L}-0.55B\textsubscript{C}-0.2SL\textsubscript{P} and vB\textsubscript{C}=-0.9P\textsubscript{L}+B\textsubscript{C}-0.18SL\textsubscript{P} to maintain visibility of the charge sensor. In the right panel we focus on the (1,0)-(0,1) transition. The charge sensor is optimized to distinguish the interdot transition. Here the virtual gate definition is set to vP\textsubscript{L}=P\textsubscript{L}-0.58B\textsubscript{C}-0.18SL\textsubscript{P} and vB\textsubscript{C}=-0.95P\textsubscript{L}+B\textsubscript{C}-0.14SL\textsubscript{P}. The gate voltages at the center of the \AI{right panel} are P\textsubscript{L}=\SI{-1381}{mV} and B\textsubscript{C}=\SI{-183}{mV}.\textbf{ b.} Schematic depicting the double occupation under P\textsubscript{L} while P\textsubscript{R} is kept below the accumulation voltage.  }
    \label{fig:fig3_virtualisation}
\end{figure*}

The charge sensor SL\textsubscript{P} effectively detects the charge state beneath the plunger P\textsubscript{L}. We begin by accumulating under P\textsubscript{L}, while keeping P\textsubscript{R} depleted, in order to avoid a lateral double quantum dot signature. Using P\textsubscript{L} and B\textsubscript{C}, we tune to a double dot regime under P\textsubscript{L}, and control the occupation of the two quantum dots QD\textsubscript{L1} and QD\textsubscript{L2}. Given their strong coupling to P\textsubscript{L},it is likely the dots are positioned underneath P\textsubscript{L}. To achieve orthogonal control of the charge occupation in the quantum dots we construct a virtual gate matrix which couples QD\textsubscript{L1} to vP\textsubscript{L}, and QD\textsubscript{L2} to vB\textsubscript{C}. This is enabled by a difference in the lever arm ratio $\alpha_{L1,BC}/\alpha_{L1,PL}<\alpha_{L2,BC}/\alpha_{L2,PL}$, where $\alpha_{D,G}$ is the lever arm between gates $G$ and quantum dot $D$. As a result, we can construct virtual gates vP\textsubscript{L} and vB\textsubscript{C} (Fig.~\ref{fig:fig3_virtualisation}) to obtain independent control of the loading onto each quantum dot, down to the single hole regime. The linearly defined virtual gate space is effective in a small voltage regime but is insufficient to virtualise subsequent transitions of the double quantum dot under P\textsubscript{L} (Fig. \ref{fig:fig3_virtualisation}a). In particular, the transitions of QD\textsubscript{L2} have a strongly varying lever arm across consecutive occupations. This difference between the quantum dots can be explained by a weaker in-plane confinement of QD\textsubscript{L2}, which is consistent with it being located in the bottom quantum well. \\

To establish that each quantum dot is indeed located in a distinct quantum well, we qualitatively estimate the location of both quantum dots. This is done by extracting the lever arm ratios of the surrounding gates to each quantum dot from the charge stability diagrams, similar to the method used by Tidjani et al. \cite{Tidjani_2023_PRApplied_VerticalDQD}. We find that the two quantum dots have approximately equal coupling to the two surrounding barrier gates B\textsubscript{L} and B\textsubscript{C}. In particular we determine $\alpha_{L1,BC}/\alpha_{L1,PL}\approx\alpha_{L1,BL}/\alpha_{L1,PL}\approx1.0$ and $\alpha_{L2,BC}/\alpha_{L2,PL}\approx\alpha_{L2,BL}/\alpha_{L2,PL}\approx1.6$ (see Supplementary~III) for the corresponding charge stability diagrams). These lever arms indicate that both quantum dots are equidistant in position between B\textsubscript{L} and B\textsubscript{C}. We note that B\textsubscript{L} and B\textsubscript{C} have similar shape and are fabricated in the same layer and we therefore ignore geometric effects. On the other hand $\alpha_{L1,SC_L}/\alpha_{L1,PL}\approx\alpha_{L2,SC_L}/\alpha_{L2,PL}\approx0.4$, indicates that neither quantum dot is significantly closer to SC\textsubscript{L}. \\

Together these findings suggest that the quantum dots are vertically stacked beneath plunger gate P\textsubscript{L}. Since the quantum dots are well-defined with a distinct interdot transition and charge signal to the sensor, we conclude that they are separated in the z-direction, with each quantum well confining one quantum dot. We assign QD\textsubscript{L2} to the bottom quantum well as its relative coupling to the barrier gates is larger than that of QD\textsubscript{L1}, which has a stronger in-plane confinement \cite{Tidjani_2023_PRApplied_VerticalDQD}. Moreover, an interdot transition $(N_{L1},N_{L2}+1)\rightarrow(N_{L1}+1,N_{L2})$ is induced by applying an increasingly negative P\textsubscript{L} voltage, indicating that QD\textsubscript{L1} is located closer to P\textsubscript{L}. The vertically coupled double quantum dot is visualised in Fig.~\ref{fig:fig3_virtualisation}b.\\

Our conclusions are further supported by our finding of comparable results for the two quantum dots QD\textsubscript{R1} and QD\textsubscript{R2} under P\textsubscript{R}, which we also tune to the (1,1) regime and where we similarly argue that each quantum dot is located in a different quantum well underneath P\textsubscript{R} (Supplementary~IV). This reproducibility bodes well for future efforts in operating larger arrays.\\

The observation of a distinct $(1,0)-(0,1)$ interdot transition line in the right panel of Fig.~\ref{fig:fig3_virtualisation}a indicates a distinct capacitative coupling between each quantum dot and SL\textsubscript{P}. This distinct capacitive coupling is encouraging, since the current heterostructure has a modest inter-layer separation, suggesting potential for further enhancement. The current ability to distinguish in which quantum well a charge is located is holds promise for vertical Pauli spin-blockade (PSB) readout. This gives perspective for the integration of a readout ancilla that can be used for PSB directly underneath or above a data qubit. This distinguishability furthermore allows to better study the inter-layer tunnel coupling itself. The control over the coupling between the quantum wells may be limited and largely predefined by their separation. Nonetheless, controlling the quantum dot occupation may serve as means to discretely change the tunnel coupling due to the varying wavefunction densities of different orbitals. The appreciable difference in the lever arms of the gates to the quantum dots furthermore suggests gate-based tunability of the inter-layer tunnel coupling and exchange interaction. An applied gate voltage could shift the quantum dots relative to one another, allowing to decrease their overlap and reducing the tunnel coupling. Alternatively, the gate voltage could influence the penetration of the wavefunction into the SiGe barrier. However, a more systematic study is needed to understand to which extent the charge occupation and tunnel couplings can be tuned independently \textit{in situ}.
\begin{figure*}
    \centering
    \includegraphics[width=\textwidth]{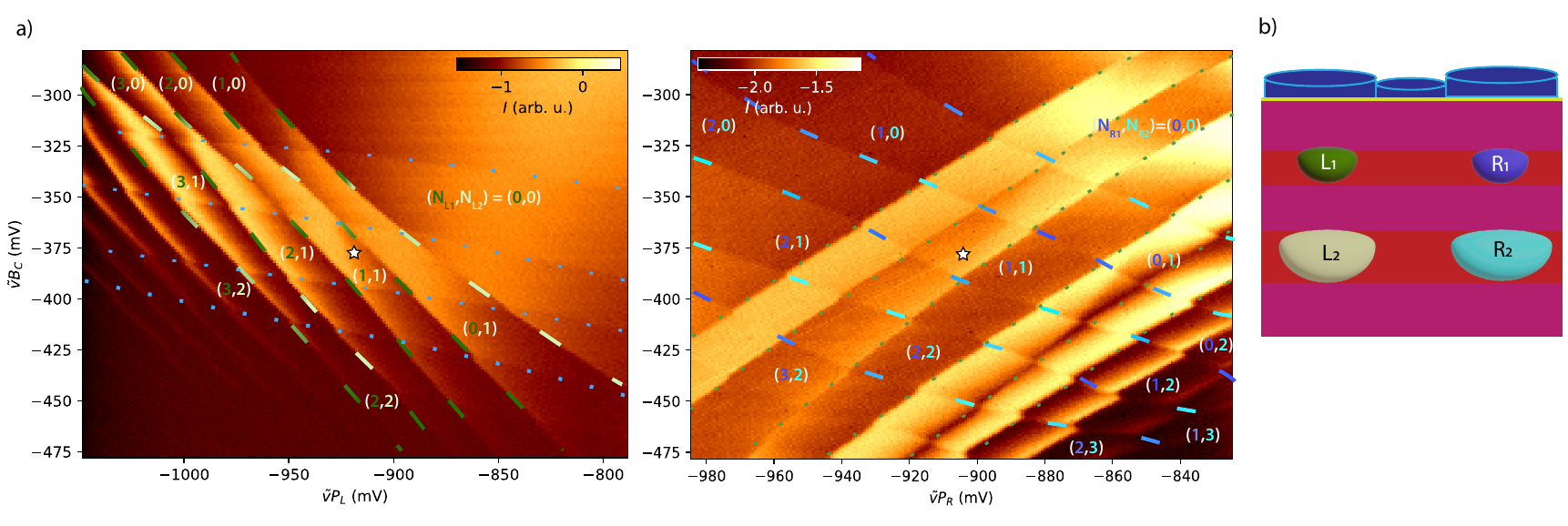}
    \caption{\textbf{Single-hole occupancy in two coupled vertical double quantum dots.} \textbf{a.} The left panel shows the charge-stability diagram with individual transitions of the double quantum dot underneath P\textsubscript{L}, where dark (light) dashed green lines correspond to reservoir transitions of QD\textsubscript{L1(2)}, serving as a guide to the eye. In addition, the blue \AI{dotted} transitions correspond to the double quantum dot under P\textsubscript{R}. We note that the individual quantum dots are poorly distinguishable due to the small lever arm differences between P\textsubscript{L} and the quantum dots underneath P\textsubscript{R}. \AI{The occupation of the top (bottom) quantum well under P\textsubscript{L} N\textsubscript{L1(2)} is indicated in the different regions}. The right panel similarly shows the charge-stability diagram with individual transitions of the double quantum dot underneath P\textsubscript{R}, with the transition to QD\textsubscript{R1(2)}indicated with dark (light) blue. The transitions corresponding to the double quantum dot under P\textsubscript{L} are indicated with a dotted green line. Again the occupation of the top (bottom) quantum wells under P\textsubscript{R} is indicated with N\textsubscript{R1(2)}. In both subfigures the virtual gate voltages are $\tilde{\text{v}}$P\textsubscript{L} = P\textsubscript{L}-0.2P\textsubscript{R}-0.17SL and $\tilde{\text{v}}$B\textsubscript{C} = B\textsubscript{C}-0.22SL and $\tilde{\text{v}}$P\textsubscript{R}=P\textsubscript{R}-0.4P\textsubscript{L}-0.5B\textsubscript{C}-0.075SL. To capture multiple transitions of the sensor in the right panel of \textbf{a}, the signal is averaged over multiple data sets at different sensor voltages SL\textsubscript{P}. The stars correspond to the same voltage values \AI{and gives the location of the (1,1,1,1) charge state}. \textbf{b.} Schematic depicting the 2x2 array. \AI{The colours match the transitions in \textbf{a}.}}
    \label{fig:fig4_1,1,1,1}
\end{figure*}

Having established individual control over the double quantum dots underneath each plunger gate, we now focus on simultaneous control over the hole occupation under both plungers to demonstrate a 2x2 array in the x-z plane. Starting in the few hole regime under P\textsubscript{R}, we maintain the (1,1) P\textsubscript{R} occupation and tune the system towards the voltage regime in which both quantum dots under P\textsubscript{L} become occupied with a single hole. The left (right) panel of Fig. \ref{fig:fig4_1,1,1,1}a demonstrates the charge-stability diagram of vP\textsubscript{L(R)} vs vB\textsubscript{C}. In each diagram one can distinguish the double quantum dot under its corresponding plunger gate, as well as additional transitions corresponding to the double quantum dot under the other plunger gate. \HT{In this figure, the upper and lower quantum dots are not virtualised with respect to each other as with in Fig. \ref{fig:fig3_virtualisation}a. in order to obtain a four quantum dot charge stability diagram while only varying two plunger gates.} In the middle of the measurement range, the vertical 2x2 array is in the (1,1,1,1) charge occupation, depicted in Fig.~\ref{fig:fig4_1,1,1,1}b. In this regime, it becomes more challenging to distinguish individual transitions from each quantum dot due to the noticeably increased interlayer tunnel coupling \AI{between QD\textsubscript{L1} and QD\textsubscript{L2} (see Supplementary~V for an analysis of the capacitive and tunnel couplings)}. This increased coupling \AI{is thought to result} from the central barrier voltage being increased to B\textsubscript{C}=13mV, compared to B\textsubscript{C}=\SI{-182}{mV} in Fig.~\ref{fig:fig3_virtualisation}, which increases the in-plane confinement. Increasing B\textsubscript{C} was necessary to achieve the desired (1,1,1,1) charge state. This high B\textsubscript{C} voltage \AI{moreover} reduces the intralayer capacitive and tunnel coupling, consistent with the observed small interdot transitions between the P\textsubscript{L} and P\textsubscript{R} quantum dots \AI{(see Supplementary~V)}.\\

In conclusion, we have established single-hole charge control over quantum dots in a double quantum well. A significant challenge remains in obtaining control over the interdot coupling and in particular when the coupling is interlayer, since the gates controlling the occupation also control the coupling. Despite this, we have shown that even in a strongly coupled system, charge sensing and orthogonal control of quantum dots in each quantum well is possible, through the construction of virtual gate matrices. Furthermore, we have demonstrated a 2x2 quantum dot array oriented perpendicular to the quantum well plane, and tuned to the (1,1,1,1) charge state. Small extensions in the system size, such as a 2x2x2 quantum dot array, may allow the study of intriguing physics arising in bilayer Hubbard models~\cite{Buterakos_2023_PRB_BilayerQSimulator}. Moreover, the ability to control single charges in multilayer systems may facilitate high-connectivity semiconductor quantum processors.
\AI{\section*{Supplementary Material}
In the supplementary material (included below) we give details on the experimental setup and the regime the charge sensor is in. We also provide data allowing to triangulate the vertical double quantum dots under P\textsubscript{L} as well as P\textsubscript{R}. Finally we analyse several anti-crossings of the charge stability diagrams to give a crude assessment of the capacitive and tunnel couplings between the quantum dots.} 

\section*{Acknowledgements}
We thank Sander de Snoo for software development and Alberto Tosato and Corentin Déprez for useful discussions.
\section*{Data Availability}
The code, analysis, and raw data supporting the findings of this study are openly available in a Zenodo repository: https://doi.org/10.5281/zenodo.10513179 
\section*{Funding}
We acknowledge support through a Dutch Research Council (NWO) Domain Science (ENW) grant and a European Research Council (ERC) Starting Grant QUIST (850641).
\section*{Competing Interests}
\AI{At the time of publication A.S. is employed by Equal1 Laboratories (The Netherlands) B.V. The remaining authors declare no competing interest.}

\renewcommand{\thefigure}{S\arabic{figure}}
\renewcommand{\thetable}{S\arabic{table}}
\renewcommand{\theequation}{S\arabic{equation}}

\makeatletter
\def\@email#1#2{%
 \endgroup
 \patchcmd{\titleblock@produce}
  {\frontmatter@RRAPformat}
  {\frontmatter@RRAPformat{\fontsize{9}{10}\selectfont\produce@RRAP{ Authors to whom the correspondence should be addressed: {#2}}}\frontmatter@RRAPformat}
  {}{}
}%
\makeatother

\onecolumngrid

\title{Supplementary Material: Coupled vertical double quantum dots at single-hole occupancy} 

\newpage

\section{Supplementary Information: Experimental Setup}
The measurements were performed in a Bluefors LD400 dilution refrigerator, with a nominal base temperature of \SI{10}{mK}. The gate voltage was applied using Qblox QCM AWG modules, with \SI{6}{dB} attenuation at the \SI{50}{K} and \SI{4}{K} plates. The DC current through the charge sensor was measured using a Keithley DMM 6500 digital multimeter.

\AI{\section{Supplementary Information: Charge sensor tuning}}

\begin{figure}[h!]
    \centering
    \includegraphics[width=0.7\textwidth]{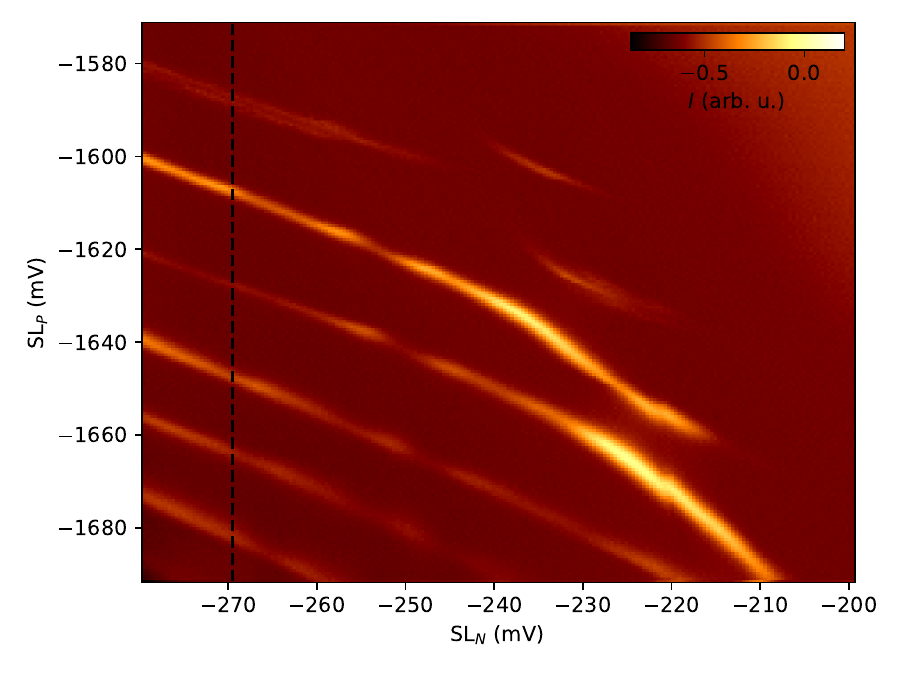}
    \caption{\AI{\textbf{Charge stability diagram of charge sensor.} The current through the charge sensor is measured as function of SL\textsubscript{N} and SL\textsubscript{P}. The linecut in Fig. 1c of the main manuscript is taken at the dashed vertical line.}}
    \label{fig:Sensor_CSD}
\end{figure}

\AI{Optimal operation of the charge sensor occurs when the Coulomb peaks are well defined. In a bilayer device this may be complicated by the possible existence of an interacting double quantum dot beneath the charge sensor. To avoid this we tune to a regime where only one of the quantum wells contributes to transport as indicated by the signature of a single quantum dot (dashed line in Fig. \ref{fig:Sensor_CSD}).} 

\section{Supplementary Information: Lever Arms for Double Quantum Dot under P\textsubscript{1}}
\label{appA:Triangulate_P1}
\begin{figure*}
    \centering
    \includegraphics[width=0.9\textwidth]{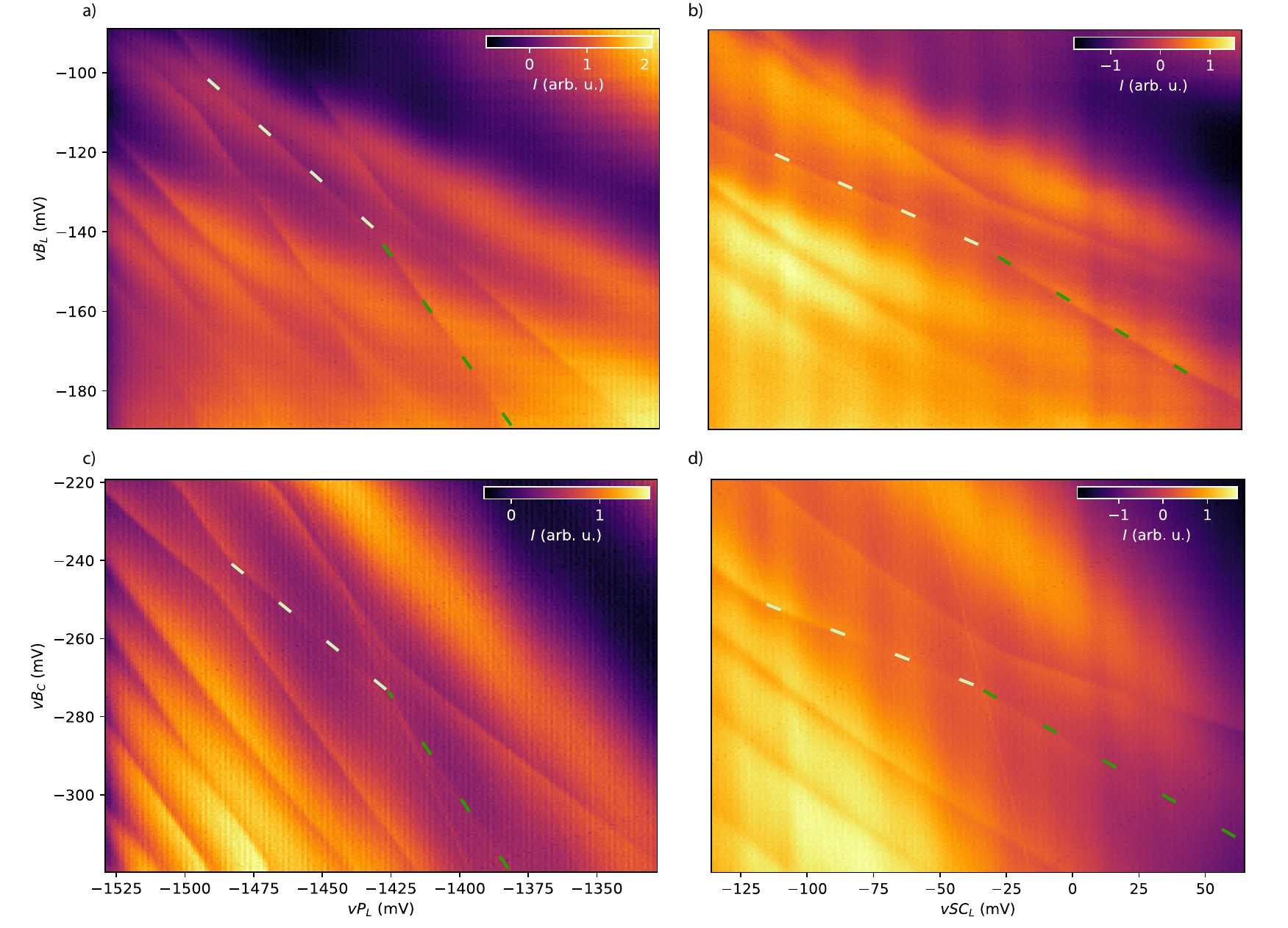}
    \caption{\textbf{Charge stability diagrams of the double quantum dot under P\textsubscript{L}.} We extract the slopes corresponding to the loading of quantum dot L\textsubscript{1(2)} indicated in the dark (light) green to retrieve the lever arm ratios of the gates to the quantum dots. These are the reservoir transitions corresponding to the loading onto the (1,1). During these measurements the virtual gates are defined to maintain high charge sensor visibility: vP\textsubscript{L}=P\textsubscript{L}-0.06SL\textsubscript{P}, vSC\textsubscript{L}=SC\textsubscript{L}-0.1SL\textsubscript{P},vB\textsubscript{L}=B\textsubscript{L}-0.4SL\textsubscript{P} and vB\textsubscript{C} = B\textsubscript{C}-0.04SL\textsubscript{P}. \AI{Note that the virtualisation here is different from the figures in the main text}. The gate voltage values at the centre of these datasets are P\textsubscript{L} = \SI{-1331}{mV}, B\textsubscript{C} = \SI{-203}{mV}, SC\textsubscript{L} = \SI{-36}{mV} and B\textsubscript{L} = \SI{-139}{mV}.}
    \label{app:fig:triangulate_left}
\end{figure*}

We triangulate the position of the double quantum dots most strongly coupled to P\textsubscript{L}, whose charge stability diagram is in Fig.~2 of the main text. This is obtained from the lever arms ratios $\alpha_{d,G_1}/\alpha_{d,G_2}$ where $\alpha_{d,G}$ is the lever arm between gates $G$ and dot $d$. These lever arm ratios are extracted from the slopes of the transition lines in the charge stability diagrams (Fig. \ref{app:fig:triangulate_left}). While subsequent reservoir transitions have different lever arms, in our analysis we only consider the reservoir transitions corresponding to the loading into the (1,1) state. To determine the lever arm ratio $\alpha_{d,P_L}/\alpha_{d,SC_L}$ for any quantum dot $d$ we combine the lever arms $\alpha_{d,B_C}/\alpha_{d,SC_L}$ and $\alpha_{d,P_L}/\alpha_{d,B_C}$. We justify this approach based on the very similar slopes for the reservoir transitions in the charge stability diagram of P\textsubscript{L} and SC\textsubscript{L}, which doesn't allow us to directly identify a slope $\alpha_{d,P_L}/\alpha_{d,SC_L}$ with a particular quantum dot $d$.\\\\
The extracted lever arm ratios are summarized in table \ref{app:tab:lever_arms_P1}. From the observation that each quantum dot couples similarly to the barrier gates on either side, and neither quantum dot couples dominantly to the screening gate SC\textsubscript{L}, we conclude that the quantum dots are centred around the same point in the x-y plane. Given that the quantum dots are sufficiently distinct and don't effectively merge into a single quantum dot, they are understood to be in distinct quantum wells, with quantum dot L\textsubscript{1(2)} being in the top (bottom) well for the reasons outlined in the main text.

We note that in this analysis the compensation on the charge sensor SL\textsubscript{P} is neglected, which we warrant through its minor effect on the quantum dot compared to the plunger and barrier gates.

\begin{table}
\caption{\textbf{Lever arm ratios for the quantum dots under P\textsubscript{L}.} The lever arm ratios for quantum dot L\textsubscript{1(2)} are extracted from the slope of the dark (light) green line in the corresponding charge stability diagrams in Fig. \ref{app:fig:triangulate_left}. The compensation on the charge sensor SL\textsubscript{P} has been neglected in this analysis. $\alpha_{SC_L}/\alpha_{P_L}$ is calculated by combining the other ratios, since the transitions of the individual quantum dots could not be distinguished from this data set. An error of 3mV is assumed when determining the slope.}
\begin{tabular}{|l|l|l|l|l|l|}
\hline
QD & $\alpha_{B_L}/\alpha_{P_L}$ & $\alpha_{B_C}/\alpha_{P_L}$ & $\alpha_{B_L}/\alpha_{SC_L}$ & $\alpha_{B_C}/\alpha_{SC_L}$ & $\alpha_{SC_L}/\alpha_{P_L}$ \\ \hline
L1   &  $0.97\pm0.08$  & $1.04\pm0.09$  & $2.45\pm0.19$ &  $2.45\pm0.20$ & $0.41\pm0.04$ \\ \hline
L2   & $1.57\pm0.15$ & $1.59\pm0.17$ & $3.48\pm0.38$ & $3.74\pm0.55$ &  $0.44\pm0.05$ \\ \hline
\end{tabular}
\label{app:tab:lever_arms_P1}
\end{table}
\newpage
\section{Supplementary Information: Double Quantum Dot under P\textsubscript{R}}
\label{appB:Triangulate_P2}
\begin{figure}
    \centering
    \includegraphics[width=1\textwidth]{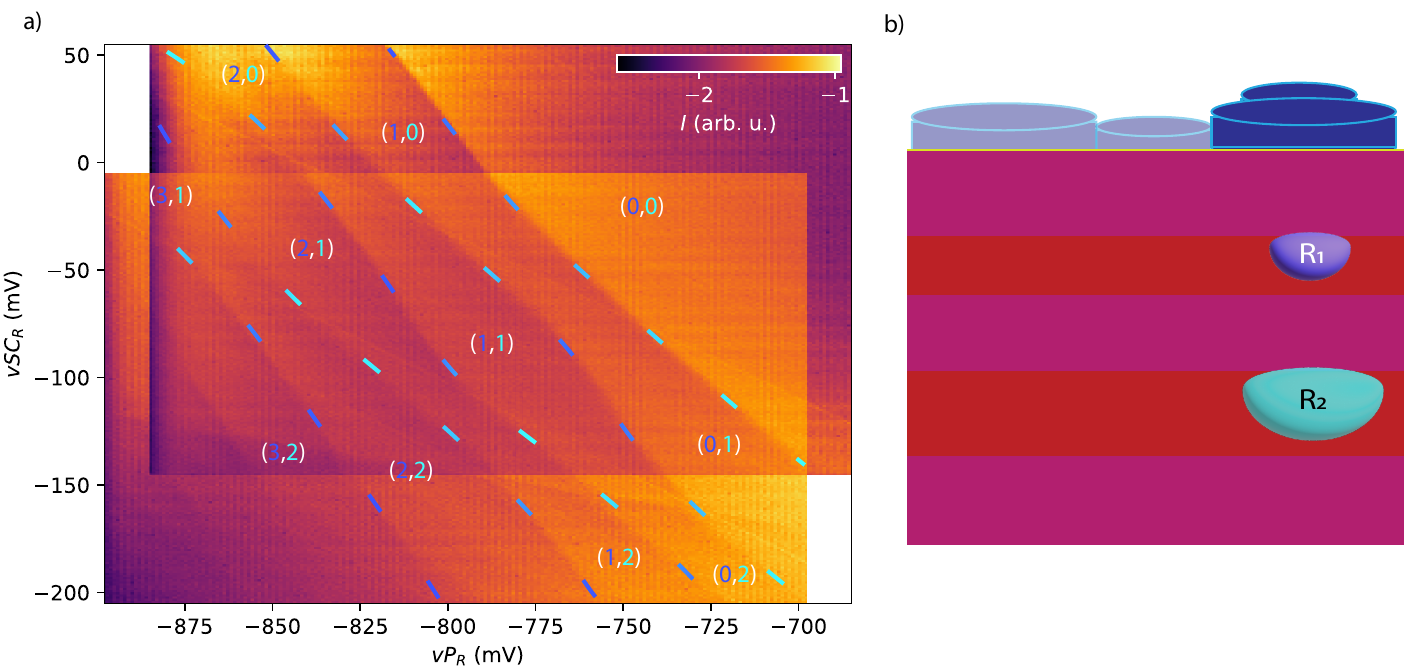}
    \caption{\textbf{A double quantum dot under P\textsubscript{R} in the few hole regime}. Two overlapping data sets demonstrate multiple transitions of distinct quantum dots that are coupled to plunger gate P\textsubscript{2}. The occupation (N\textsubscript{R1},N\textsubscript{R2}) is denoted in the different charge regions. Here vP\textsubscript{R} = P\textsubscript{R}-0.055SL\textsubscript{P} and vSC\textsubscript{R} = SC\textsubscript{R}-0.075SL\textsubscript{P}. We attribute the charge transition crossing the y-axis at about \SI{-120}{mV} to a spurious dot near SC\textsubscript{R}. \AI{Note that the virtualisation here is different from the figures in the main text}. The gate voltage values at the centre of these data sets are given by P\textsubscript{R} = \SI{-791}{mV} and SC\textsubscript{R} = \SI{-75}{mV}. The two data-sets that are used are averaged at the points of overlap. The dashed lines are added to guide the eye.}
    \label{app:fig:DoubleDotP2}
\end{figure}

\begin{figure*}
    \centering
    \includegraphics[width=\textwidth]{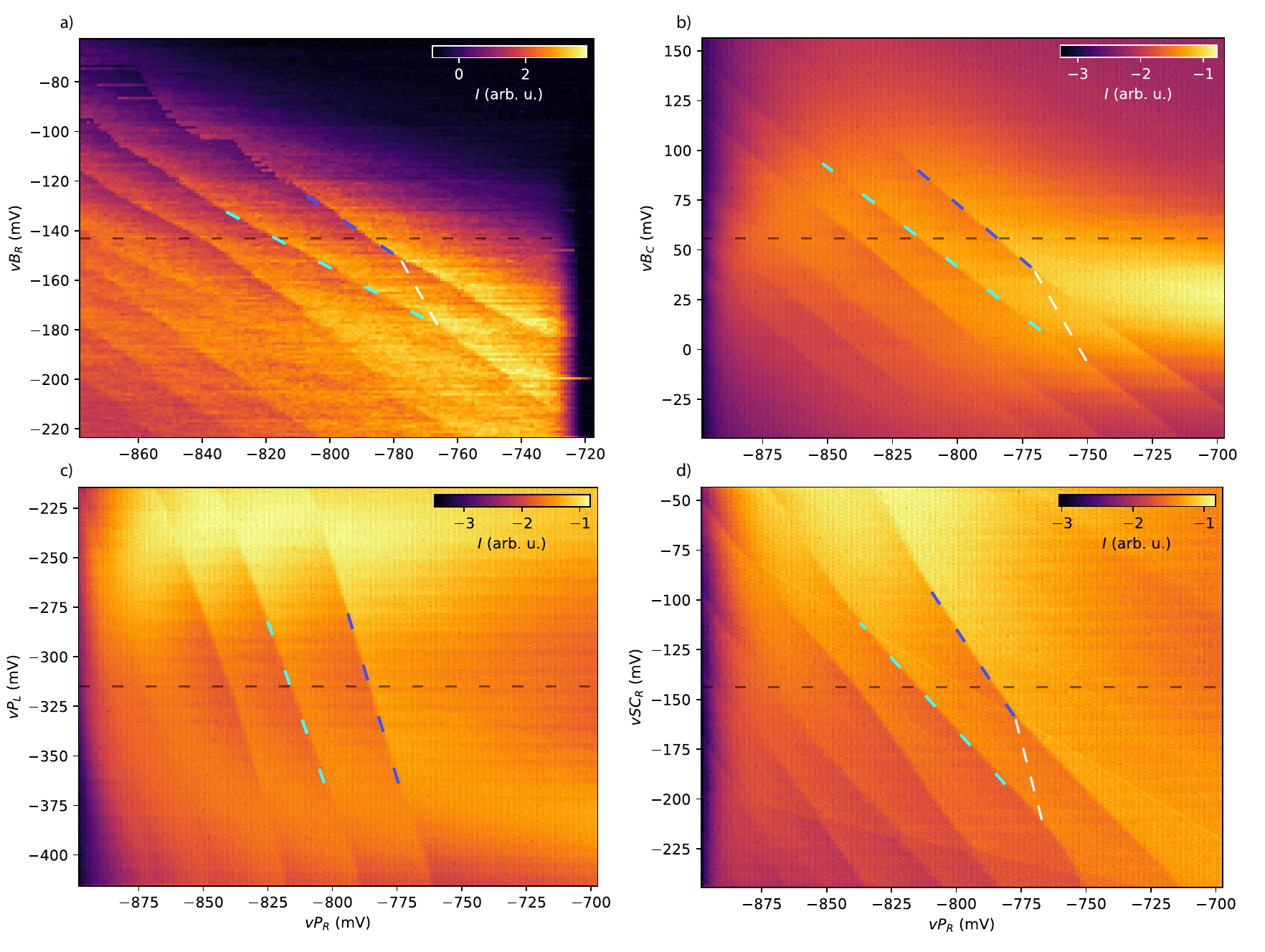}
    \caption{\textbf{Charge Stability Diagrams of the double dot under P\textsubscript{R}.} We extract the slopes corresponding to the loading of quantum dot QD\textsubscript{R1(2)} indicated in the dark (light) blue to retrieve
the lever arm ratios between the gates and the quantum dots. The black dashed lines indicate points of equal voltage across the different CSDs. The white line connects the two triple points at the (0,1)-(1,0) interdot transition, whenever these are distinguishable. The CSD with B\textsubscript{R} was taken using a slower scan due to the limited bandwidth of the DC-line connected to B\textsubscript{R}. In that CSD the latching effect becomes pronounced as B\textsubscript{R} is more positive, as the quantum dots are loaded from the right reservoir. Across this data, the virtual gates are defined such that the sensor-voltage is compensated as such: vP\textsubscript{R} = P\textsubscript{R}-0.055SL\textsubscript{P}, vSC\textsubscript{R}=SC\textsubscript{R}-0.075SL\textsubscript{P}, vB\textsubscript{C}=B\textsubscript{C}-0.2SL\textsubscript{P}, vP\textsubscript{L}=P\textsubscript{L}-0.1SL\textsubscript{P} and vB\textsubscript{R}=B\textsubscript{R}-0.045SL\textsubscript{P}. The exception to this is CSD subfigure \textbf{a} which has vP\textsubscript{R}=P\textsubscript{R}-0.05SL\textsubscript{P}. \AI{Note that the virtualisation here is different from the figures in the main text}. The gate voltage values at the centre of these datasets are given by P\textsubscript{R} = \SI{-798}{mV},P\textsubscript{L} = \SI{-315}{mV}, B\textsubscript{C} = \SI{56}{mV}, SC\textsubscript{R} = \SI{-144}{mV} and B\textsubscript{R} = \SI{-143}{mV}.
}
    \label{app:fig:triangulate_right}
\end{figure*}

We demonstrate the formation of the double quantum dots under P\textsubscript{R} and triangulate their position as we have done in supplementary \ref{appA:Triangulate_P1}. For these measurements, the charge sensor on the left side of the device is used. The increased distance between the P\textsubscript{R} and the sensor results in the weaker signal in Fig. \ref{app:fig:DoubleDotP2} compared to Fig.~2 of the main text. Still, we distinguish two distinct transitions lines, each attributed to a different dot. The lack of further transitions in the (0,0) region shows that indeed the single-hole regime is reached.  Plunger gate P\textsubscript{L} is depleted such that no dots underneath it are occupied.
\\\\
We further triangulate the exact positions of the double dots by determining the lever arm ratios of the surrounding gates to these quantum dots (fig. \ref{app:fig:triangulate_right}) and summarize the values in Table \ref{app:tab:lever_arms_P2}). The reservoir transitions used for extracting the lever arms have been denoted with blue and cyan dashed lines in Fig.~\ref{app:fig:triangulate_right}, as these transitions can be consistently identified across the different charge stability diagrams. We see that the lever arm ratios are not as homogeneous as was found for the quantum dots under P\textsubscript{L}, as in particular quantum dot QD\textsubscript{R2} seem to be coupled more with B\textsubscript{R} than B\textsubscript{C}. Still neither quantum dot couples particularly weakly or strongly to any surrounding gate, and therefore they are unlikely to be spurious dots underneath any particular gate, as that would result in strong coupling to that gate and low coupling to a further-positioned gate. Based on the studied reservoir transitions alone we can not decisively argue that the two quantum dots must be located in different quantum wells. In particular, based on the given lever arm ratios (tab. \ref{app:fig:triangulate_right}), an alternative interpretation would be that both quantum dots are located in the same quantum well, both between barrier gates B\textsubscript{C} and B\textsubscript{R}, but with QD\textsubscript{R2} closer towards SC\textsubscript{R} than QD\textsubscript{R1}. However, this interpretation suggests that an increasingly negative voltage on SC\textsubscript{R} would be able to transfer a hole from QD\textsubscript{R1} into QD\textsubscript{R2}. Yet the (0,1)-(1,0) interdot transition (white dashed line in fig. \ref{app:fig:triangulate_right}) suggests that an increasingly negative voltage on SC\textsubscript{R} (B\textsubscript{C}, B\textsubscript{R} or P\textsubscript{R}) would localise the single hole into QD\textsubscript{R1}. This would suggest that $\alpha_{R1,G}>\alpha_{R2,G}$ with G being SC\textsubscript{R},B\textsubscript{C},B\textsubscript{R} or P\textsubscript{R} which, together with the lever arm ratio in table \ref{app:tab:lever_arms_P2}, conflicts with any configuration of in-plane double quantum dots. Hence we conclude that the two quantum dots are located in different quantum wells. More precisely, by a similar reasoning as for the double quantum dots under P\textsubscript{L} we suggest that QD\textsubscript{R1(2)} is located in the top (bottom) quantum well (see Fig.~\ref{app:fig:DoubleDotP2}b for a schematic).

\begin{table}
\caption{\textbf{Lever arm ratios for the quantum dots under P\textsubscript{R}.} Similar to table \ref{app:tab:lever_arms_P1} the lever arm ratios for quantum dot QD\textsubscript{R1(2)} are extracted from the slope of the dark (light) blue line in the corresponding charge stability diagrams in Fig.~\ref{app:fig:triangulate_right}. The compensation on the charge sensor SL\textsubscript{P} has been neglected in this analysis. An error of \SI{3}{mV} is assumed when determining the slope.}
\begin{tabular}{|l|l|l|l|l|}
\hline
QD & $\alpha_{B_R}/\alpha_{P_R}$ & $\alpha_{B_C}/\alpha_{P_R}$ & $\alpha_{P_L}/\alpha_{P_R}$ & $\alpha_{SC_R}/\alpha_{P_R}$  \\ \hline
R1   &  $1.07\pm0.16$  & $0.89\pm0.07$  & $0.23\pm0.03$ &  $0.51\pm0.05$  \\ \hline
R2   & $1.44\pm0.13$ & $1.02\pm0.06$ & $0.27\pm0.04$ & $0.69\pm0.05$  \\ \hline
\end{tabular}
\label{app:tab:lever_arms_P2}
\end{table}
\newpage
\AI{\section{Supplementary Information: Analysis of capacitive and tunnel coupling}}
\AI{The focus of this work is on the establishment of few-hole occupation in bilayer quantum dot arrays. However, the capacitive and tunnel coupling between neighbouring quantum dots are important parameters for charge and spin-manipulation\cite{Hanson2007SpinsDots,VanDerWiel2003}. While further investigations are needed to reliably extract and predict the tunability of the tunnel coupling in bilayer systems, here we provide a crude assessment of the observed couplings in the present experiments.} \\

\AI{We use the measured charge stability diagrams to evaluate these parameters through the bending of the charge transition lines. We detect the individual charge transitions using built-in scipy\cite{2020SciPy-NMeth} and scikit\cite{pedregosa2011scikit} functions, the code for which are available together with the raw data on Zenodo. The voltage values $(V_x,V_y)$ of these transitions are fitted to a two-level model\cite{Hanson2007SpinsDots, VanDerWiel2003,Huttel_2005_PRB_ChargeBending,PieroLadiere_2005_PRB_ChargeBending}:} \begin{equation}
    V_y = \begin{bmatrix}
    a_1 (V_x-x_0)+y_0 +(-)E_m/2 & t_c\\t_c & -a_2 (V_x-x_0)+y_0 +(-) E_m/2
\end{bmatrix}\label{eq:TwoLevelModel}\end{equation}
\AI{where the positive (negative) eigenvalues of this matrix give the upper (lower) branch of the charge anti-crossing, with capacitive coupling of $E_m$, tunnel coupling $t_c$, $a_{1,2}$ determining the lever arm ratio to the different quantum dots, and $x_0,y_0$ the offsets in voltage space. We note that this model assumes a constant lever-arm between the gates and both quantum dots across the voltage space which seems sufficient for the charge transitions we analyse. \\}

\AI{To estimate the capacitive and tunnel couplings we assume the lever arm $\alpha_{QD_1,P}\approx\SI{0.09}{eV/V}$ between the plunger gates and the quantum dot in the top quantum well\cite{Johnbichromatic}. This lever arm has been experimentally determined in a monolayer device with a similar plunger gate size whose quantum well is at the same depth as the top quantum well in this study. While the lever arm might deviate in our bilayer design, we believe that any corrections will be comparatively small, as confirmed by a 2D Schr\"odinger-Possion simulations predicting a similar lever arm of $\alpha\approx0.12$\cite{Tidjani_2023_PRApplied_VerticalDQD}. We will further validate this lever arm by using it to extract the charging energy of the top quantum dots and compare that with the charging energy $E_C=\SI{619}{GHz}$ of the earlier monolayer work\cite{Johnbichromatic}.\\}

\begin{figure}
    \centering
    \includegraphics[width=\textwidth]{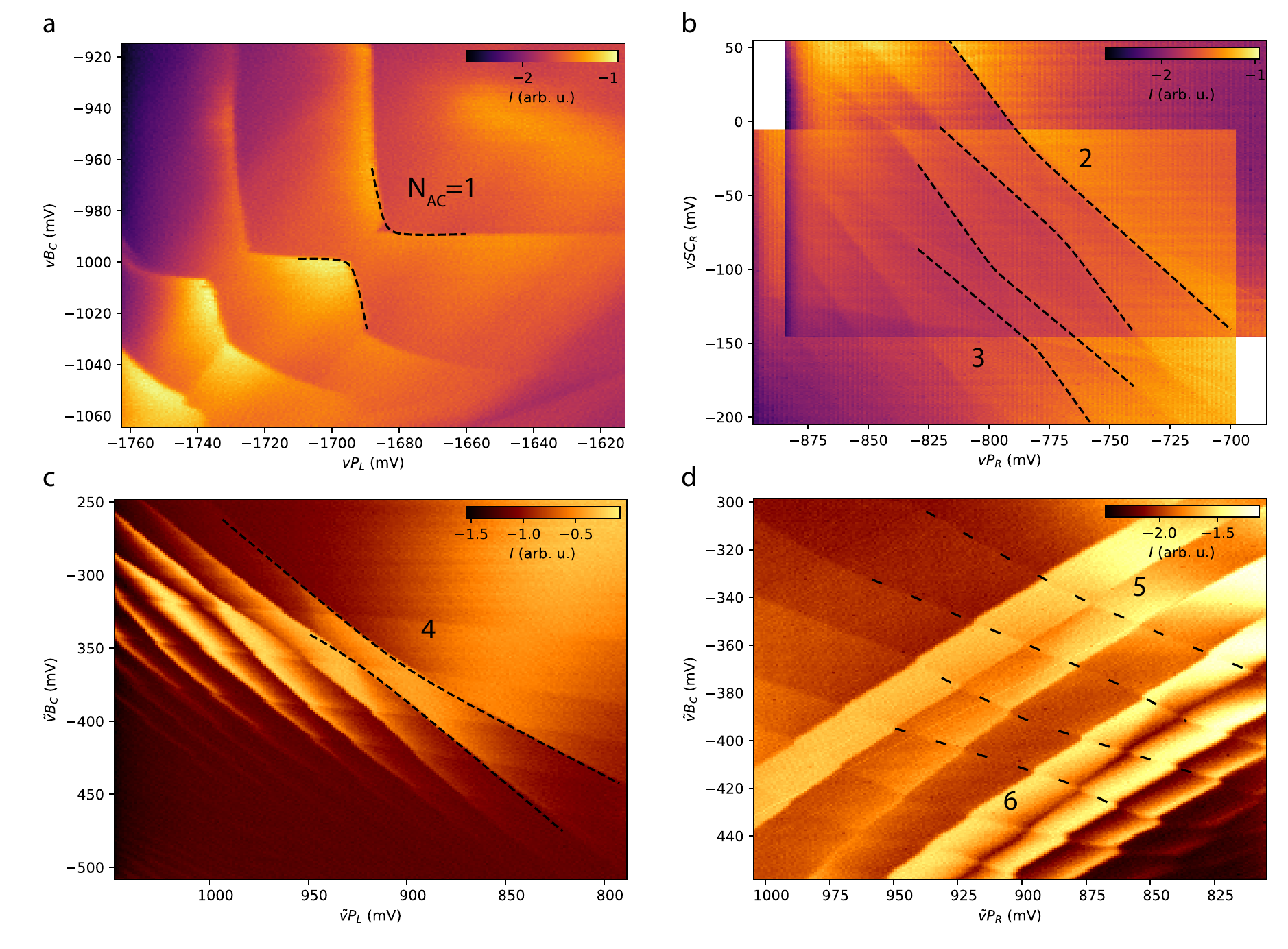}
    \caption{\AI{\textbf{Fitted charge stability diagrams to extract the vertical tunnel and capacitive couplings.} The charge stability diagrams are the same as the ones used in earlier figures, with \textbf{a}(\textbf{b}) showing the single hole regime of QD\textsubscript{L(R)} and \textbf{c}(\textbf{d}) showing the single hole regime of the coupled vertical double quantum dot pair, focusing on QD\textsubscript{L(R)}. Note that the virtualisation of the plunger and barrier gates is different across the regimes with $\tilde{\text{v}}$P\textsubscript{L} = P\textsubscript{L}-0.5B\textsubscript{C}-0.2SL\textsubscript{P}, $\tilde{\text{v}}$B\textsubscript{C} = 0.9P\textsubscript{L}+B\textsubscript{C}-0.18SL\textsubscript{P}, $\tilde{\text{v}}$P\textsubscript{R} = P\textsubscript{R}-0.055SL\textsubscript{P}, $\tilde{\text{v}}$SC\textsubscript{R} = SC\textsubscript{R}-0.075SL\textsubscript{P}, $\tilde{v}$P\textsubscript{L} = P\textsubscript{L}-0.2P\textsubscript{R}-0.17SL\textsubscript{P}, $\tilde{\text{v}}$B\textsubscript{C} = B\textsubscript{C}-0.22SL\textsubscript{P}, $\tilde{\text{v}}$P\textsubscript{R} = P\textsubscript{R}-0.4P\textsubscript{L}-0.5B\textsubscript{C}-0.075SL\textsubscript{P}. The dashed lines are fits of the charge transitions according to eigenvalues of the model in equation \ref{eq:TwoLevelModel}. Based on the lever arm $\alpha_{QD_1,P}=\SI{0.09}{eV/V}$ we extract a charging energy of the top quantum dot in subfigure \textbf{a.} of $E_C\approx416\pm10\SI{}{GHz}$, in \textbf{b.} of $E_C\approx718\pm22\SI{}{GHz}$, in \textbf{c.} of $E_C\approx699\pm20\SI{}{GHz}$ and in \textbf{d.} of $E_C\approx583\pm8\SI{}{GHz}$, where the uncertainties are based on a measurement uncertainty of 1mV. We note that this is comparable with single quantum well devices with a similar depth of the quantum well and comparable gate geometry\cite{Johnbichromatic}, which had $E_C\approx\SI{619}{GHz}$.}}
    \label{supp:fig:verticalDotFits}
\end{figure}

\begin{figure}
    \centering
        \includegraphics[width=0.6\textwidth]{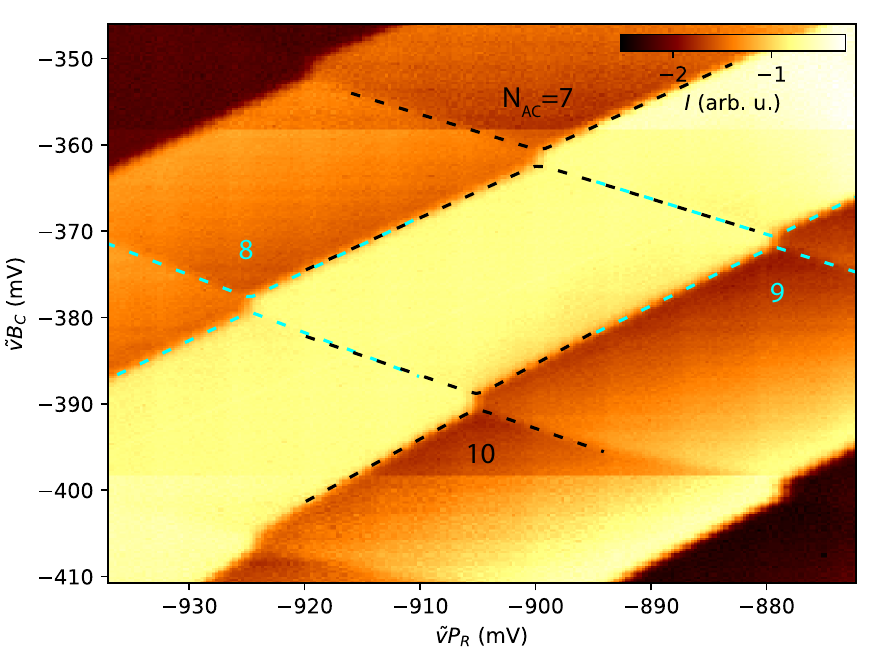}
    \caption{\AI{\textbf{Fitted charge stability diagrams to extract the lateral tunnel and capacitive couplings.} The data is a higher-resolution scan of the right panel in Fig.~3a of the main text. Dashed lines correspond to fits of the four indicated anti-crossings. We note that charge state between the anti-crossings corresponds to the (1,1,1,1) regime.}} 
    \label{supp:fig:lateralDotFits}
\end{figure}

\AI{The tunnel coupling and capacitive coupling are extracted based on the fits displayed in Fig.~\ref{supp:fig:verticalDotFits}, together with the knowledge of the lever arm ratios from tables \ref{app:tab:lever_arms_P1} and \ref{app:tab:lever_arms_P2}. The results can be found in table \ref{supp:tab:Couplings}. 
With the exception of the vertical tunnel coupling of the left double quantum dot pair in the (1,1,1,1) regime, the extracted tunnel couplings are comparable to the targeted tunnel coupling of $\SI{16}{GHz}$ at this quantum well separation\cite{Tidjani_2023_PRApplied_VerticalDQD}. When both double quantum dots are occupied, we note a significant increase in the vertical capacitive coupling as well as an increase of the tunnel coupling for the left double quantum dot. This is explained by an increased in-plane confinement due to the barrier gate voltage $B_C$ from $\SI{-182}{mV}$ in Fig.~\ref{supp:fig:verticalDotFits}a to $\SI{13}{mV}$ in Fig.~\ref{supp:fig:verticalDotFits}c,d. It should however be stressed that the uncertainty of the fit is considerable, and to draw precise conclusions future investigation is needed to collect data on the tunability of these couplings over a larger voltage space.\\\\
Additionally, the lateral couplings between the two double quantum dots have been analysed around the (1,1,1,1) charge state. As before, the two-level model (eq. \ref{eq:TwoLevelModel}) has been fitted to the charge transitions, with the resulting fits found in Fig.~\ref{supp:fig:lateralDotFits}, with the extracted couplings in table~\ref{supp:tab:LateralCouplings}. For these lateral transitions, it is not well determined between which quantum dot the coupling occurs, since at some transition the vertical double quantum dots hybridize. For simplicity, we again assume a lever arm of $\alpha=\SI{0.09}{eV/V}$. Clearly, there is a diminished but non-zero capacitive coupling. This decrease in lateral capacitive coupling is to be expected given the larger distance between the quantum dots, as well as the planar configuration.  The tunnel coupling seems to be diminished as well but to a lesser extent. Due to the uncertainty of the method, the results are not conclusive. In future work, one might consider performing photon-assisted tunnelling to determine it directly~\cite{VanDerWiel2003}. In general, we note that typically the lateral tunnel coupling will not decrease with the same order as the capacitive coupling since the tunnel barrier of the SiGe buffer is of a different magnitude than the in-plane confinement, and the tunnel coupling depends exponentially on the barrier height and width. }

\begin{table}[]
\caption{\AI{\textbf{Extracted vertical tunnel and capacitive couplings.} Based on model \ref{eq:TwoLevelModel} fitted to the different anti-crossings N\textsubscript{AC} in Fig.~\ref{supp:fig:verticalDotFits} we extract the tunnel and capacitive couplings in the vertical direction. To extract the energies we assume a lever arm $\alpha_{QD1,P}=\SI{0.09}{eV/V}$ between the plunger gate and the quantum dot in the top well. The transitions corresponding to the steeper slopes are assumed to correspond to the quantum dot in the upper quantum well, as justified in the main text. To account for the virtualisation factors, tables \ref{app:tab:lever_arms_P1} and \ref{app:tab:lever_arms_P2} are used. The uncertainties are standard deviations based on the first-order derivative near the optimum of the least-square optimizer.}}
\begin{tabular}{|l|l|l|l|l|l|l|}
\hline
N\textsubscript{AC}            & 1 & 2 & 3 & 4 & 5 & 6 \\ \hline
$t_c$ (GHz) & $10\pm18$   & $21\pm6$   & $10\pm7$   & $83\pm22$   & $11\pm5$   &  $5\pm6$  \\ \hline
$E_m$ (GHz) & $100\pm18$ &  $280\pm40$  & $116\pm30$   & $293\pm102$   &  $331\pm34$  &  $260\pm30$  \\ \hline
\end{tabular}
\label{supp:tab:Couplings}
\end{table}

\newpage

\begin{table}[h!]
\caption{\AI{\textbf{Extracted lateral tunnel and capacitive couplings.} Based on model \ref{eq:TwoLevelModel} fitted to the different anti-crossings N\textsubscript{AC} in Fig.~\ref{supp:fig:lateralDotFits} we extract the tunnel and capacitive couplings in the lateral and possibly diagonal direction. To extract the energies we assume a lever arm $\alpha=\SI{0.09}{eV/V}$ between the plunger gate and the quantum dot underneath it, independent of the quantum well. To account for virtualisation, tables \ref{app:tab:lever_arms_P1} and \ref{app:tab:lever_arms_P2} are used. The uncertainties are standard deviations based on the first-order derivative near the optimum of the least-square optimizer. Note that because for N\textsubscript{AC} the estimated $t_c$ approaches 0, the first derivative of the optimizer blows up, here $E_m$ is taken as the standard deviation.}}
\begin{tabular}{|l|l|l|l|l|}
\hline
N\textsubscript{AC}            & 7 & 8 & 9 & 10 \\ \hline
$t_c$ (GHz) & $4\pm4$   & $2\pm6$   & $0\pm E_m$   & $4\pm3$   \\ \hline
$E_m$ (GHz) & $33\pm3$ &  $25\pm3$  & $23\pm2$   & $28\pm2$  \\ \hline
\end{tabular}
\label{supp:tab:LateralCouplings}
\end{table}
\clearpage
%

\end{document}